\documentclass{article}

\usepackage{arxiv}

\usepackage[utf8]{inputenc} % allow utf-8 input
\usepackage[T1]{fontenc}    % use 8-bit T1 fonts
\usepackage{hyperref}       % hyperlinks
\usepackage{url}            % simple URL typesetting
\usepackage{booktabs}       % professional-quality tables
\usepackage{amsfonts, amsmath}       % blackboard math symbols
\usepackage{nicefrac}       % compact symbols for 1/2, etc.
\usepackage{microtype}      % microtypography
\usepackage{graphicx}
\usepackage{natbib}
\usepackage{array, multirow}
\usepackage{doi}
\usepackage[version=4]{mhchem}

\title{Hybrid Quantum-Classical Inverse Design of Metasurfaces for Tailored Narrow Band Absorption}

\author{ \href{https://orcid.org/0000-0002-7211-0773}{\includegraphics[scale=0.06]{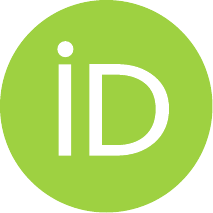}\hspace{1mm}Sreeraj Rajan Warrier} \\
	Department of Physics\\
	Mahindra University\\
	Hyderabad, India 500043 \\
	\texttt{rajan21pphy004@mahindrauniversity.edu.in} \\
	\And
	\href{https://orcid.org/0000-0003-3428-0170}{\includegraphics[scale=0.06]{orcid.pdf}\hspace{1mm}Jayasri Dontabhaktuni} \\
	Department of Physics\\
	Mahindra University\\
	Hyderabad, India 500043 \\
	\texttt{jayasri.d@mahindrauniversity.edu.in} \\
}

\date{}

\renewcommand{\undertitle}{}
\renewcommand{\shorttitle}{\textit{arXiv} Template}

\begin{document}
\maketitle

\begin{abstract}
The inverse design of metasurfaces poses a considerable challenge because of the intricate interdependencies that exist between structural characteristics and electromagnetic responses.  Traditional optimization methods require significant computational resources and frequently do not produce the most effective solutions. This study presents a hybrid quantum-classical machine learning approach known as Latent Style-based Quantum GAN (LaSt-QGAN). This method integrates a Variational Autoencoder (VAE) with a Quantum Generative Adversarial Network (QGAN) to enhance the optimization of metasurface designs aimed at achieving narrow-band absorption and unidirectionality. The proposed method results in a reduction of training time by 10X and a decrease in data requirements by 40X when compared to traditional GAN-based approaches. The produced metasurface designs demonstrate a high fidelity in relation to the target absorption spectra compared with the classical GAN. Additionally, the integration of a material look-up table facilitates manufacturability by allowing for the substitution of predicted material properties with viable alternatives, all while preserving performance accuracy. Moreover the model is able to generate Q-factor upto the order of $10^4$, while the training dataset has Q-factor upto the order of $10^3$.
\end{abstract}

% keywords can be removed

\section{Introduction}
The integration of material characteristics with nanoscale structures is a significant problem in modern material discovery and photonics research, crucial for achieving the intended functionality. Metasurfaces, two-dimensional structures adept at modulating electromagnetic waves in novel ways, have attracted significant attention due to rapid advancements in optics and photonics~\cite{yu2014flat}. The surfaces in consideration have unique optical properties, including narrow-band absorption and unidirectionality, which have enabled their widespread use in sensing, imaging, and telecommunications~\cite{kildishev2013planar}. However, the design of metasurfaces to meet specific functional requirements remains challenging due to the non-unique solutions and the complex interplay between their structural properties and electromagnetic responses~\cite{genevet2017recent}. Traditional methods for metasurface design often rely on extensive numerical optimization or iterative trial-and-error techniques~\cite{dasdemir2023computational}. While these approaches can provide usable designs, they frequently demand substantial computing resources and may not reliably produce the most effective solutions~\cite{molesky2018inverse}.

In addressing these issues, the emergence of inverse design has been noted, utilizing advanced algorithms to systematically determine the structural features necessary for achieving the desired optical properties~\cite{jiang2019global}. In recent years, the integration of artificial intelligence and machine learning into inverse design methodologies has significantly advanced the optimization processes associated with metasurface design. References include~\cite{liu2018training, So2020}. Recent research efforts have explored a variety of computational intelligence models and deep learning architectures for the purposes of material and structural optimization~\cite{sanchez2018inverse, tu2020machine, bingman2020deep}. The utilization of these approaches within the domain of photonics has been demonstrated, particularly in the development of core-shell nanoparticles and multi-layer thin films, through the application of one-dimensional tandem networks~\cite{liu2018training, so2019acs}.

The implementation of generative algorithms, particularly Generative Adversarial Networks (GANs), presents a significant advantage in the field. Generative Adversarial Networks (GANs) represent significant progress in the field of generative learning models, with extensive applications in domains including image processing, video analysis, and molecular design~\cite{ian2015adversarial}. The models presented in this study are formulated by means of the training of encoded images which incorporate material along with structural parameters. This approach facilitates the creation of metasurface designs that align with designated target absorption spectra~\cite{Yeung2021global}.  Despite these advancements, the intricate high-dimensional parameter spaces and complex design landscapes associated with metasurfaces continue to present major obstacles for conventional computational techniques. For instance, the occurrence of mode collapse frequently observed in Generative Adversarial Networks (GANs) remains a critical issue~\cite{Kossale2022mode, cerezo2022challenges}. The computational requirements inherent in Generative Adversarial Networks (GANs) are approaching the limits defined by Moore's law. The process of generating images with dimensions of \(512 \times 512\) pixels through the application of Generative Adversarial Networks (GANs) requires the handling of 158 million parameters. The procedure involves a training period of two days, utilizing a dataset that consists of 14 million instances, and is carried out on a setup of 512 tensor processing units (TPUs)~\cite{deng2019unsupervised}.

The implementation of Quantum Generative Adversarial Networks (QGANs) in the realm of quantum computing offers a novel approach to the inverse design of metasurfaces, indicating significant potential for transformation in this field. They exhibit considerable potential in improving the design process through their ability to efficiently traverse and optimize parameter spaces~\cite{preskill2018quantum, lloyd2018quantum}. The integration of quantum computing with the use of GANs facilitates a more rapid and efficient convergence towards optimal solutions, thereby advancing the development of precise and innovative metasurface designs across multiple technological domains~\cite{dallaire2018quantum}. Contemporary theoretical studies suggest that quantum generative algorithms may provide an exponential edge over classical algorithms, consequently generating increased interest in the theoretical and experimental aspects of quantum GANs~\cite{Huang2021, Tsang2023, Stein2021, Romero2021, chang2024latent}.

The main aim of Generative Adversarial Networks (GANs) is to replicate the essential features of the training dataset~\cite{ian2015adversarial}. This methodology involves the simultaneous training of two neural networks, with one serving the role of a discriminator and the other acting as a generator.  The generator's primary role is to produce fake information that accurately reflects the attributes of the genuine training dataset. In contrast, the discriminator is responsible for discerning the differences between original and fake information. The collaborative interactions inherent within these networks contribute to an enhancement of their performance, facilitated by competitive interactions occurring all through the training procedure. The generator aims to delineate the distribution of samples obtained from an unknown data source, referred to as \(P_{\text{data}}\), as demonstrated by the training dataset.  Beginning with a foundational latent distribution, represented as \(P_z\), the generator performs a transformation to produce \(P_g = G(P_z)\). The aim is for \(P_g\) to achieve convergence with \(P_{\text{data}}\), a condition that is rarely satisfied beyond the simplest cases. This study introduces a conditional vector \(\gamma\) that encompasses the absorption spectra in conjunction with unidirectional data. The discriminator (\(D\)) and generator (\(G\)) engage in a minimax game, in which the generator aims to minimize the likelihood that the discriminator can correctly differentiate between real and synthetic data, while the discriminator seeks to maximize this likelihood. The value function pertinent to this game is delineated as follows~\cite{Yeung2021global}:

\begin{equation}
\label{eq1}
\min _G \max _D V(D, G) = \mathbb{E}_{\boldsymbol{x} \longrightarrow p_{\text{data}}}[\log D(\boldsymbol{x, \gamma})] + \mathbb{E}_{\boldsymbol{z} \longrightarrow p_z}[\log (1 - D(G(\boldsymbol{z, \gamma})))]
\end{equation}

\begin{itemize}
\item \(\mathbb{E}\) : Expected value
\item \(x\): Real data sample
\item \(z\): Random Noise
\item \(\gamma\): Conditional vector (Absorption Spectra)
\item \(D(x)\): Probability that the discriminator accurately identifies real data as authentic
\item \(G(z)\): Generated data
\item \(D(G(z))\): Probability that the discriminator categorizes generated data as authentic
\end{itemize}

The neural networks engage in a process of iterative training, wherein each network is dedicated to the minimization of a specific loss function.

\begin{equation}
\label{eq2}
\begin{aligned}
\mathcal{L}_D &= -[y \cdot \log (D(x, \gamma)) + (1 - y) \cdot \log (1 - D(G(z, \gamma)))], \\
\mathcal{L}_G &= -[(1 - y) \cdot \log (D(G(z, \gamma)))]
\end{aligned}
\end{equation}

Within this framework, the variable \(y\) serves as an indicator for binary classification, where a value of \(y = 1\) signifies authentic data, and a value of \(y = 0\) indicates synthetic data.

The proposed formulation of \(\mathcal{L}_G\) represents a departure from the original setup of Generative Adversarial Networks (GANs), which aims to minimize \(\log[1 - D(G(z, \gamma))]\).  This specific methodology has been recognized for producing insufficient gradients~\cite{ian2015adversarial, inkawhich_dcgan}.

This research presents a novel hybrid classical-quantum Generative Adversarial Network (QGAN) approach, referred to as Latent Style-based Quantum GAN (LaSt-QGAN)~\cite{chang2024latent}. This study presents a framework that integrates two separate components: a pretrained variational autoencoder (VAE) and a patch quantum GAN~\cite{Kingma2013AutoEncodingVB, Huang2021}. This research introduces a methodology that combines the benefits of rapid computational techniques with the flexible design characteristics found in image-based deep learning frameworks, while also addressing the common issue of mode collapse that is often associated with Generative Adversarial Networks (GANs).  This approach, which is grounded in quantum generative learning as it pertains to imagery, effectively combines the predictability of material properties and structural components with improved computational training and flexible design approaches.  This study specifically outlines the material and structural properties, such as refractive indices and plasma frequencies, in conjunction with their dimensional parameters (e.g., meta-atom thicknesses) and environmental factors. These elements are categorized into three distinct channels corresponding to color images, as evidenced by prior research~\cite{Yeung2021global}.

\section{Methods} 

\begin{figure}[ht]
\centering
\includegraphics[height=0.5\linewidth, width=\linewidth]{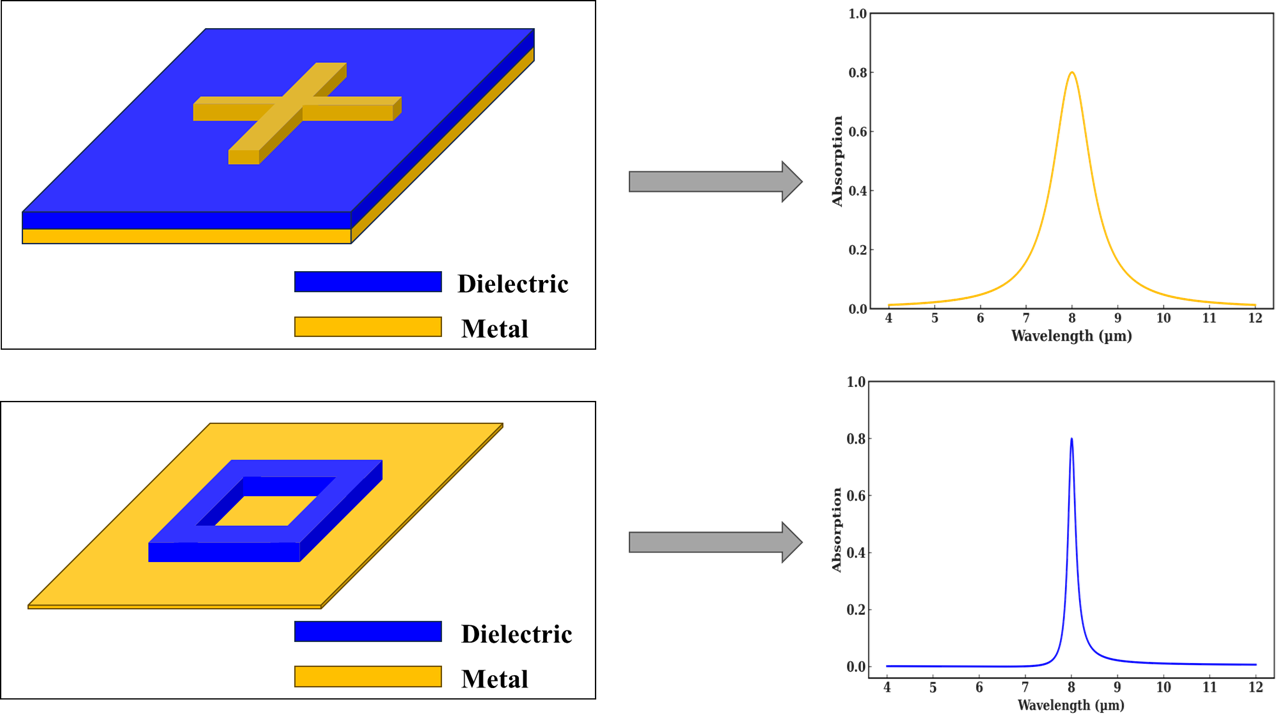}
\caption{Metal-insulator-metal (MIM) and hybrid dielectric metasurface unit cells demonstrate distinct absorption spectra characterized by Lorentzian and Fano shapes, correspondingly.}
\label{fig:1}
\end{figure}

This research investigates the inverse design methodology, focusing on two specific types of absorbing metasurfaces, as depicted in Figure~\ref{fig:1}. This study examines two distinct categories of absorbing metasurfaces: hybrid dielectric metasurfaces and metal-insulator-metal (MIM) configurations. The configuration of metal-insulator-metal (MIM) structures is characterized by the presence of a dielectric layer situated between two metallic components. The examined structures exhibit a detailed Lorentzian-shaped absorption response, making them particularly effective for tasks concerning thermal radiation and energy harvesting~\cite{Liu2011, Neutens2009}. Hybrid dielectric metasurfaces utilizing a dielectric resonator over a metallic film substrate exhibit cavity effects that facilitate the generation of an asymmetric, narrow-band Fano resonance. This characteristic renders them especially advantageous for sophisticated functions in the field of photonics~\cite{Chen2019}.

\begin{figure}[ht]
\centering
\includegraphics[height=0.5\linewidth, width=\linewidth]{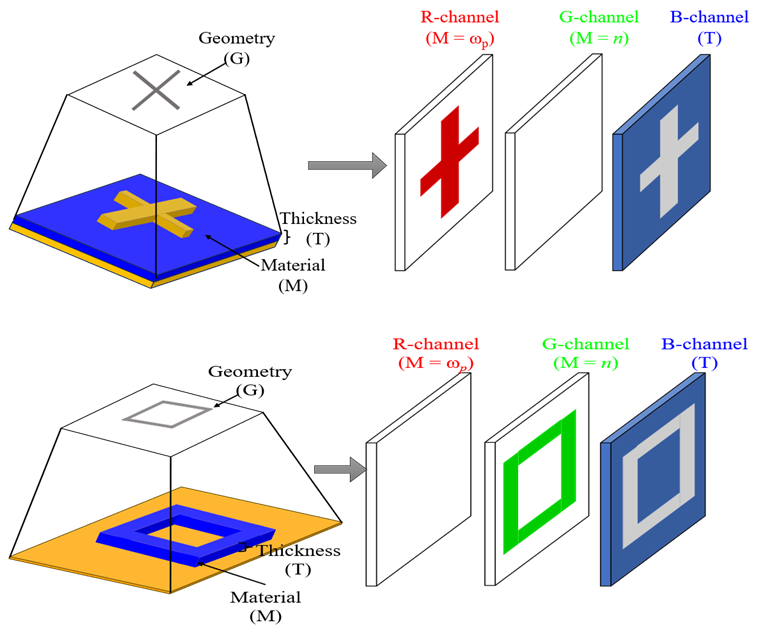}
\caption{RGB encoding of metasurface unit cells, where geometry is embedded and material properties, thickness, and structural features are mapped to individual color channels.}
\label{fig:2}
\end{figure}

The encoding approach illustrated in Figure~\ref{fig:2} begins with the collection of planar topologies (G), material properties (M), and dielectric layer thicknesses (T) relevant to both MIM as well as hybrid dielectric metasurfaces. The parameters being examined are incorporated into the three separate channels of a colored image. The red channel is specifically associated with the plasma characteristics and the geometric configuration present within metal-insulator-metal (MIM) structures. The green channel demonstrates a correlation with the refractive index as well as the geometrical arrangement of dielectric elements within hybrid structures. The evaluation of the dielectric layer's thickness is conducted via the blue channel for both categories. In the context of MIM structures, the encoding produces a red-blue color spectrum, whereas for hybrid dielectric structures, it generates a green-blue color spectrum.

The dataset utilized for the absorption spectra training consists of 500 distinct metasurface unit cell layouts, represented as image-vector combinations. The structures presented in this study were derived from seven unique shape types, as outlined in the research conducted by Yeung et al 2021~\cite{Yeung2021global, raman2024multiclass}. The investigation into the modeling of MIM and hybrid dielectric structures is performed utilizing unit cells with dimensions of \(3.2 \times 3.2\) and \(7.5 \times 7.5\) \(\mu\text{m}^2\), respectively. The restricted availability of qubits poses a substantial constraint on our capacity to encode the complete set of values related to absorption spectra within the framework of quantum circuits. The parameters designated to represent the conditional vector are derived subsequent to the fitting of the Fano equation, specifically Resonance frequency(\(\nu_0\)), Linewidth (\( \Gamma\)), peak absorption (\(A_0\)), and Fano asymmetry parameter (\(q\))~\cite{limonov2017fano}. The parameters are articulated through a conditional vector denoted as (\(\gamma\)), which is further augmented by the inclusion of random noise represented as (\(z\)). The design was transformed into a three-dimensional pixel representation, defined as a \(64 \times 64 \times 3\) pixel "RGB" image, as illustrated in Figure \ref{fig:2}. Finite element method simulations were performed on the generated designs using COMSOL Multiphysics to acquire an absorption spectrum vector range spanning from 4 to 12~\(\mu\)m~\cite{comsol}.

\begin{figure}[ht]
\centering
\includegraphics[height=0.68\linewidth, width=\linewidth]{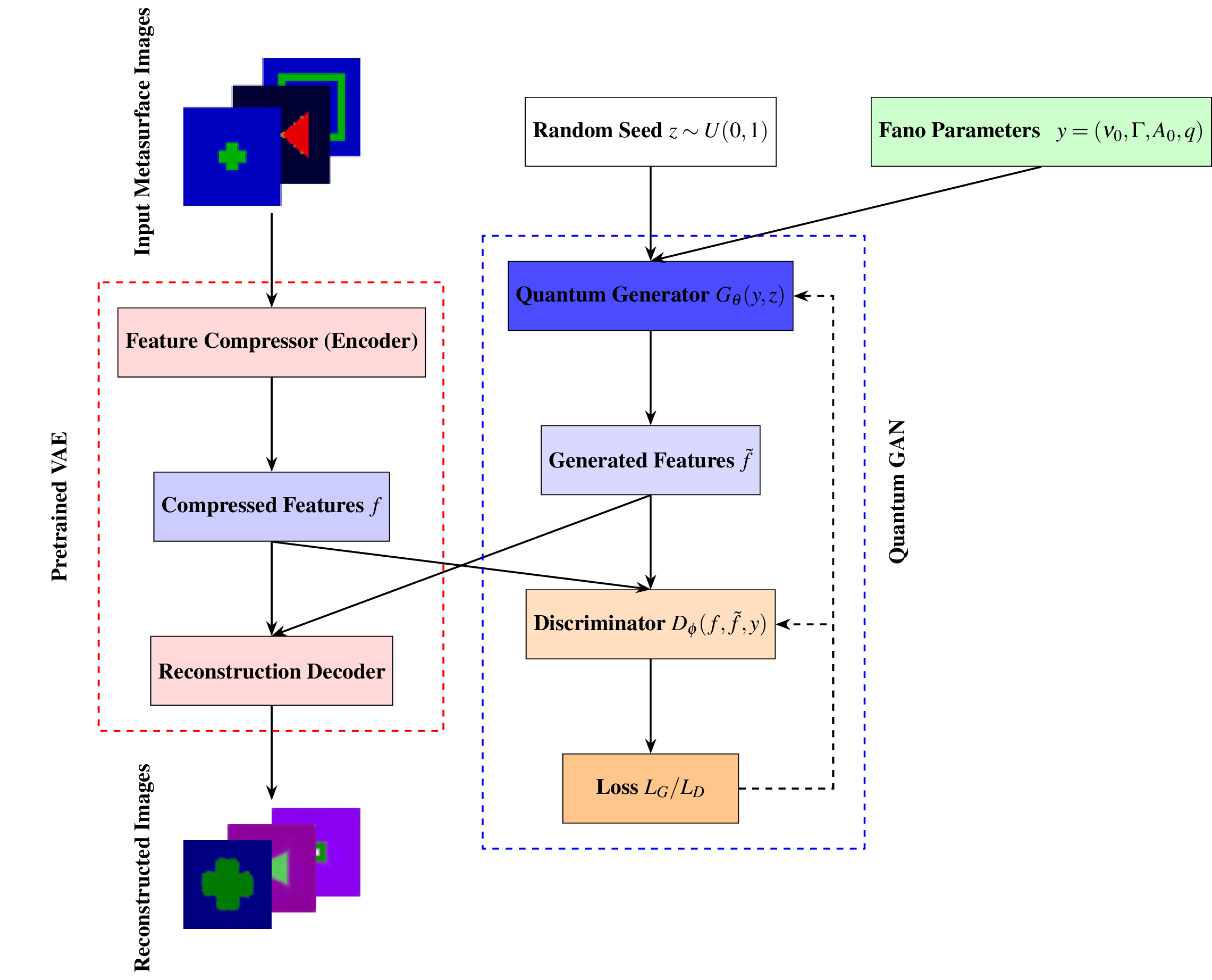}
\caption{Schematic Training Architecture of a Hybrid Quantum Generative Adversarial Network (QGAN) with a Pretrained Variational Autoencoder (VAE)}
\label{fig:3}
\end{figure}

In the color-encoding phase for absorption metasurfaces, the plasma resonance frequencies, as defined by the Drude model to represent metallic resonators, were utilized to encode the red channel. The plasma frequencies observed in this study are quantified as follows: for gold (\ce{Au}), the plasma frequency is measured at \(\omega_P = 1.91\) PHz; for silver (\ce{Ag}), it is \(\omega_P = 2.32\) PHz; and for aluminum (\ce{Al}), the plasma frequency is recorded at \(\omega_P = 3.57\) PHz. The refractive indices of the dielectric resonators were measured for the green channel, yielding values of \(n = 2.4\) for zinc selenide (\ce{ZnSe}), \(n = 3.46\) for silicon (\ce{Si}), and \(n = 4\) for germanium (\ce{Ge}).  The optical constants pertinent to the mid-infrared wavelength spectrum employed in the modeling were utilized to derive the material features. The blue channel encapsulated the dielectric thickness, demonstrating a variation range between 100 and 950 nm. In order to maintain the integrity of the "RGB" color system, a normalization process was applied to all encoded values, ensuring they fell within the specified range of 0 to 255.

\begin{figure}[ht]
\centering
\includegraphics[width=\linewidth, height=0.3\linewidth]{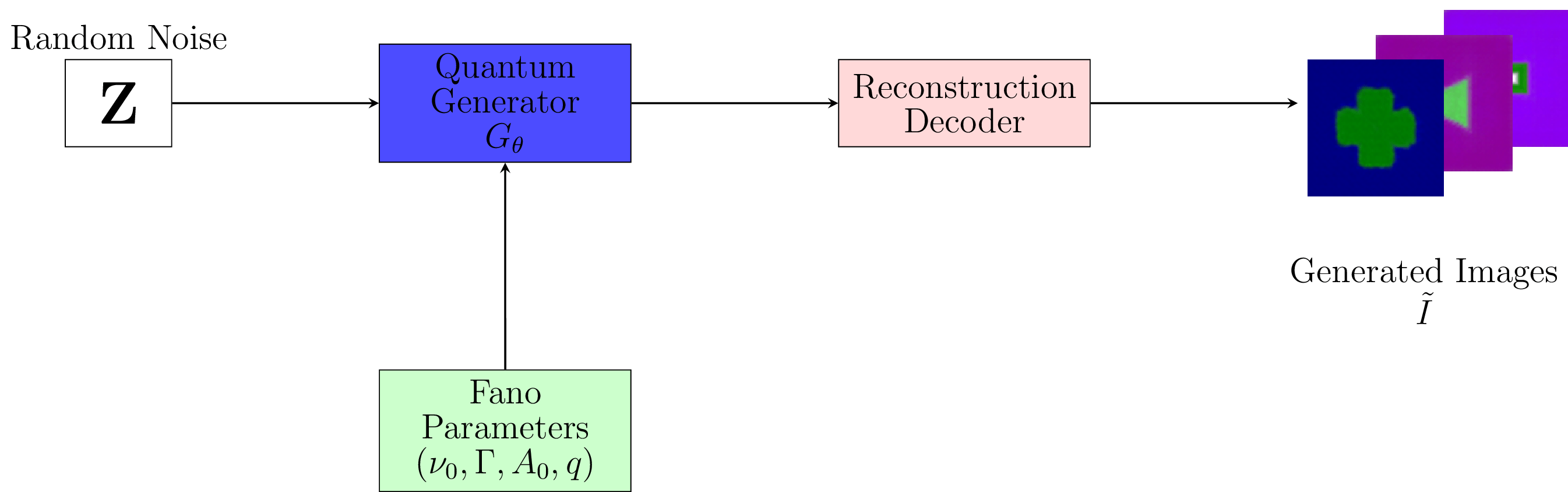}
\caption{Testing and Validation Process of a Hybrid Quantum GAN: From Latent Space Sampling to the Generation of Synthetic Metasurface Structure Images}
\label{fig:4}
\end{figure}

The comprehensive workflow for both training and inference processes associated with LaSt-QGAN is visually represented in Figures~\ref{fig:3} and~\ref{fig:4}, respectively. Figure~\ref{fig:3} illustrates the continuous feedback-driven process in which the discriminator and generator engage in a non-cooperative game aimed at enhancing the general effectiveness of LaSt-QGAN. In the course of the training procedure, images provided undergo transformation via Variational Autoencoders (VAE) to effectively encode essential data in a dense format, thereby accommodating the constraints of resource-limited Noisy Intermediate-Scale Quantum (NISQ) computers, rather than engaging in a pixel-by-pixel learning approach. Figure~\ref{fig:4} illustrates the application of the decoder from the pretrained Variational Autoencoder (VAE) in generating images of the metasurface structures. This occurs during the operation of the fully-trained quantum generators subsequent to the training of the LaSt-QGAN. The Python script was executed on a GPU server, specifically the NVIDIA DGX-1, which is equipped with 1 TB of RAM, 15 TB of SSD storage, 128 CPU cores, and a single A100 (Volta) GPU, operating under the Ubuntu 20.04 operating system.

During the initial stage, essential features represented as \(x\) are derived from real images \(I\) situated within a latent field of specified dimensionality, employing a conventional convolutional variational autoencoder (VAE). The Variational Autoencoder (VAE) undergoes the pre-training phase utilizing the original image dataset characterized by dimensions of \(64 \times 64 \times 3\) pixels. This process operates as a reversible technique for accomplishing dimensionality reduction. The derived features constitute the foundational training dataset employed in the current quantum GAN learning process.

\begin{figure}[ht]
\centering
\includegraphics[height=0.45\linewidth, width=\linewidth]{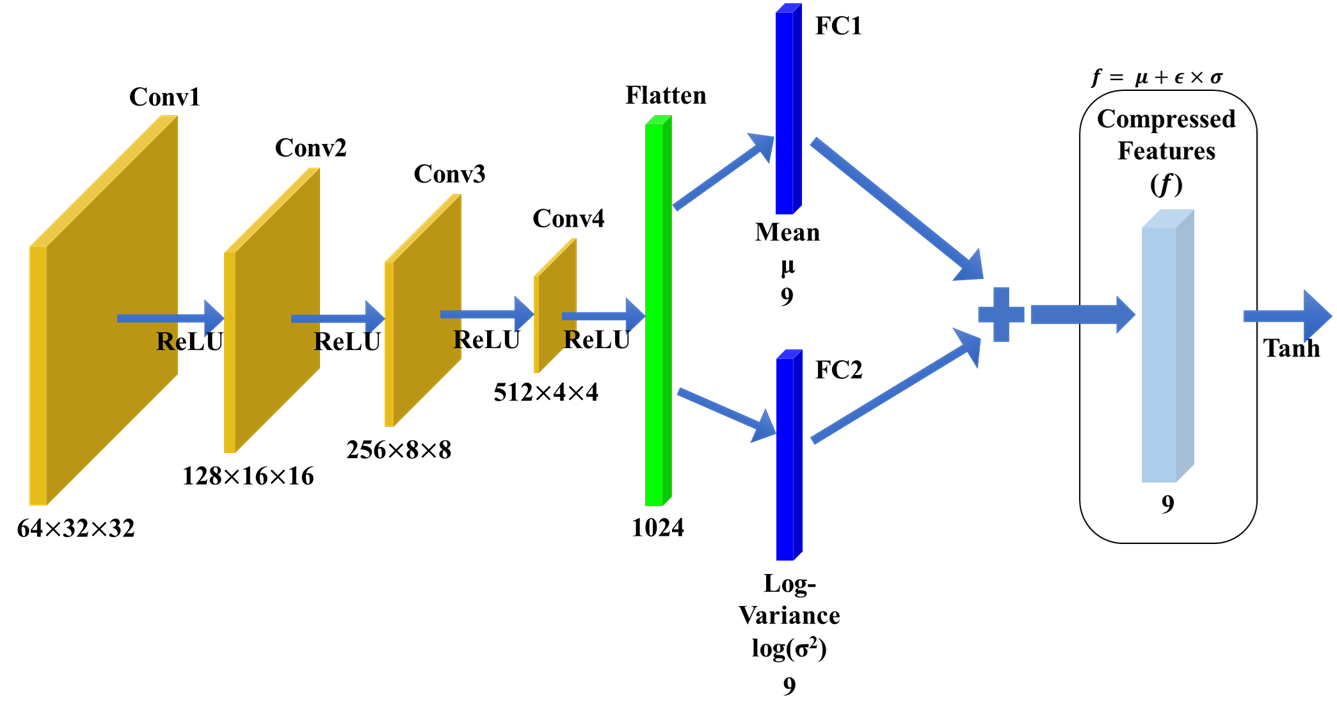}
\includegraphics[height=0.4\linewidth, width=\linewidth]{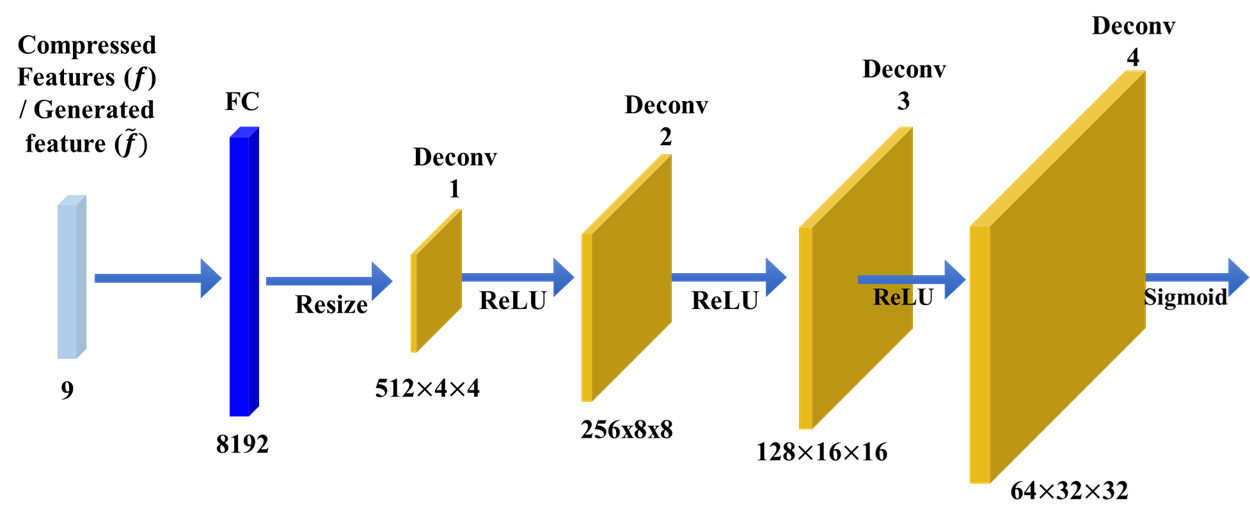}
\caption{Architecture of the $\beta$-VAE and IWAE used in the paper.}
\label{fig:5}
\end{figure}

This study presents a comparative analysis of the outcomes derived from two pretrained variational autoencoders: the $\beta$-VAE and the importance weighted autoencoders (IWAE)~\cite{higgins2017betavae, burda2015importance}. \(\beta\)-VAE and IWAE represent unsupervised neural network architectures, developed within the PyTorch framework, aimed at facilitating dimensionality reduction and data compression processes~\cite{Adam2019PyTorch}. The primary distinction between \(\beta\)-VAE and IWAE lies in their methodologies for learning the latent features of data. The \(\beta\)-Variational Autoencoder (VAE) introduces a control parameter, denoted as \(\beta\), which serves to modulate the trade-off between the fidelity of data reconstruction and the organization of hidden features within the latent space. This adjustment enhances the interpretability of the extracted features. The Importance Weighted Autoencoder (IWAE) employs a technique known as importance sampling, which enhances its ability to accurately represent intricate data patterns. In the conducted simulations, the parameter \(\beta\) was established at a value of 2 for both models, while for the Importance Weighted Autoencoder (IWAE), the number of samples was maintained at a constant value of 5. Figure~\ref{fig:5} demonstrates that both Variational Autoencoders (VAEs) function by transforming high-dimensional data into a lower-dimensional latent space and thereafter reassembling the data from these latent representations (\(x\)) through the decoder, which possesses a dimensionality comparable to that of the output generated by the quantum generator.

The quantum Generative Adversarial Network (GAN) operates as a generative framework, utilizing a quantum generator to produce synthetic attributes derived from random noise in conjunction with a conditional vector (\(\gamma\)). In parallel, a conventional discriminator is assigned the role of differentiating between genuine characteristics (\(x\)) obtained from the \(\beta\)-Variational Autoencoder (VAE) or Importance Weighted Autoencoder (IWAE) and fake features (\(\Tilde{x}\)) generated by the quantum generator, implemented via Xanadu Technology's PennyLane simulator~\cite{Bergholm2018}.

\begin{figure}[ht]
\centering
\includegraphics[height=0.55\linewidth, width=\linewidth]{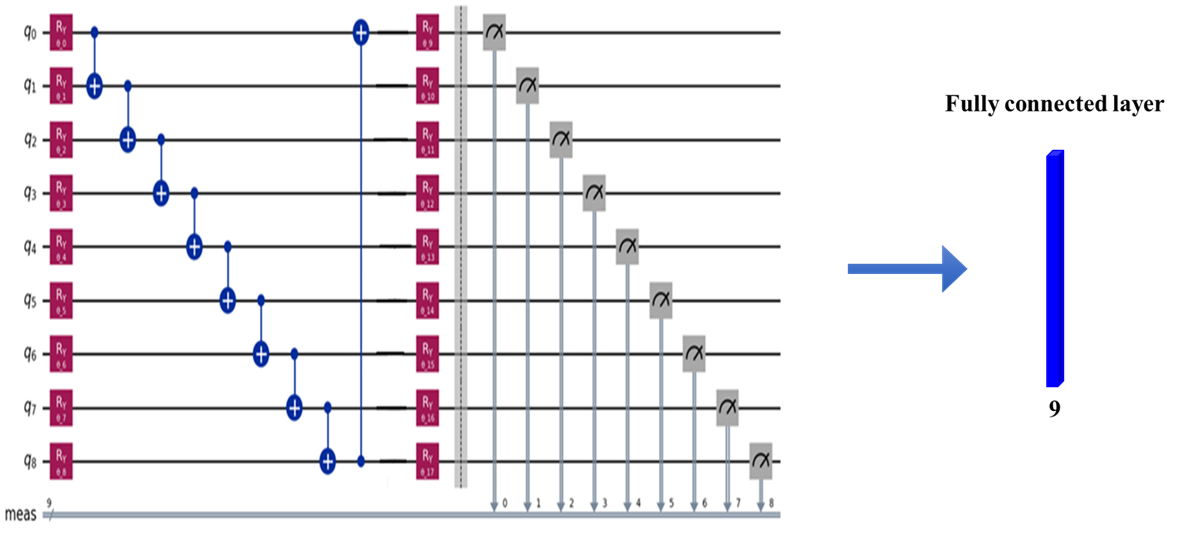}
\caption{Quantum neural network employed for the LaSt-QGAN model.}
\label{fig:6}
\end{figure}

This study presents a quantum generator, illustrated in Figure~\ref{fig:6}, which is constructed utilizing a style-based architecture~\cite{Bravo-Prieto2022}. In this framework, the rotational parameters within the training layers are influenced by latent data.  The system is designed to process input data through the concatenation of the latent noise vector (\(z\)) and conditional information (\(\gamma\)). This integration with the input (\(x\)) enhances the learning process, making it more dynamic and adaptable. The composite is subsequently channeled through a series of sub-generators, which are implemented as variational quantum circuits~\cite{cerezo2021variational, Peruzzo2014}. The continuous improvement of these sub-generators during the training process contributes to the optimization of their performance. The circuits demonstrate properties determined by adjustable variables, denoted as (\(\theta\)), which are altered throughout the training process. The current implementation features a quantum circuit structured with parameterized layers that include rotation gates identified as \(\text{RY}\). The configuration is subsequently improved by the incorporation of circular entanglement via the utilization of CNOT gates. The comprehensive unitary transformation \( U \) of the circuitry is articulated as \( U = U_i(z, \gamma) \cdot \ldots \cdot U_1(z, \gamma) \). In this formulation, each layer \( i \) of the circuit integrates an orthogonal transformation that is applied to the input \( x \).

\begin{equation}
\label{eq5}
\begin{aligned}
\theta_i &= W_i \cdot x + b_i, \\
\end{aligned}
\end{equation}

where,
\begin{itemize}
    \item \( W_i \): The orthogonal weight matrix of size \(n_{qubits} \times n_{qubits}\) with $n_{qubits}$ being the number of qubits~\cite{kerenidis2021classical}.
    \item \( b_i \): The bias vector of size \(n_{qubits}\).
    \item $x$ : [{$z|\gamma$}] is the concatenation of latent noise ($z$) and conditional vector ($\gamma$).
    \item \( \theta_i \): The rotation angles for the quantum gates in the \(i\)-th layer of the quantum circuit.
\end{itemize}

Within each sub-generator, the input (\(x\)) is incorporated into the rotation angles (\(\theta_i\)) corresponding to the quantum gates through the affine transformation delineated in equation~\ref{eq5}. The weight matrices undergo orthogonalization to enhance accuracy and efficiency while mitigating issues related to vanishing or inflating gradients in quantum neural network architectures. Furthermore, these matrices are uniformly distributed within the range of [-0.01, 0.01]~\cite{kerenidis2021classical}. The bias \(b_i\) is randomly selected from a uniform distribution that ranges from -0.01 to 0.01. The quantum circuit subsequently implements the specified rotation angles through a series of rotation gates, which are succeeded by circular entanglement layers, thereby guaranteeing the appropriate entanglement of qubits. The achievement of entanglement is facilitated by the implementation of CNOT gates, which establish a circular configuration of interconnections among qubits.

In the course of the training procedure, the model systematically optimizes the parameter set (\(\Theta = \{W_i, b_i\}\)) across all layers, iteratively adjusting the angles to enhance the generative capacity of the quantum model.  The results obtained from the quantum circuit are expressed in terms of Pauli-Z expectation values, denoted as \(\langle \sigma_z \rangle\).  The aforementioned process is systematically replicated across all sub-generators, with each one playing a significant role in the overall output. In order to maintain consistency, the latent space is constrained to the interval [-1, 1], which corresponds to the output range of \(\langle \sigma_z \rangle\) produced by the quantum generator. This architectural framework enables the representation of complex, high-dimensional information, making it suitable for a variety of tasks in the field of quantum machine learning.

This study utilizes a classical discriminator, which is a fully connected neural network, designed to distinguish between the latent space ($x$) and the synthetic feature representation (\(\Tilde{x}\)) in the context of the LaSt-QGAN. The model utilizes input features in conjunction with an equivalent conditional vector (\(\gamma\)), which are subsequently processed through a series of layers. At the outset, the conditional vector data is subjected to a linear transformation in order to align its dimensionality with that of the input features. The classical discriminator is composed of three fully connected layers. The initial layer contains 128 nodes, which is succeeded by a second layer featuring 64 nodes. Subsequent to each of these layers, a Leaky ReLU activation function is applied, characterized by a negative slope of 0.2~\cite{Jin2020ReLu}. The concluding layer consists of a single node, followed by a Sigmoid activation function~\cite{Cybenko1989}. This configuration produces a probability score that indicates the genuineness of the input data, thereby assessing whether it is real or synthetic. The feedback process plays a vital role in directing the quantum generator to enhance the realism of its output features.

\section{Results}

\begin{figure}[ht]
\centering
\includegraphics[width=\linewidth]{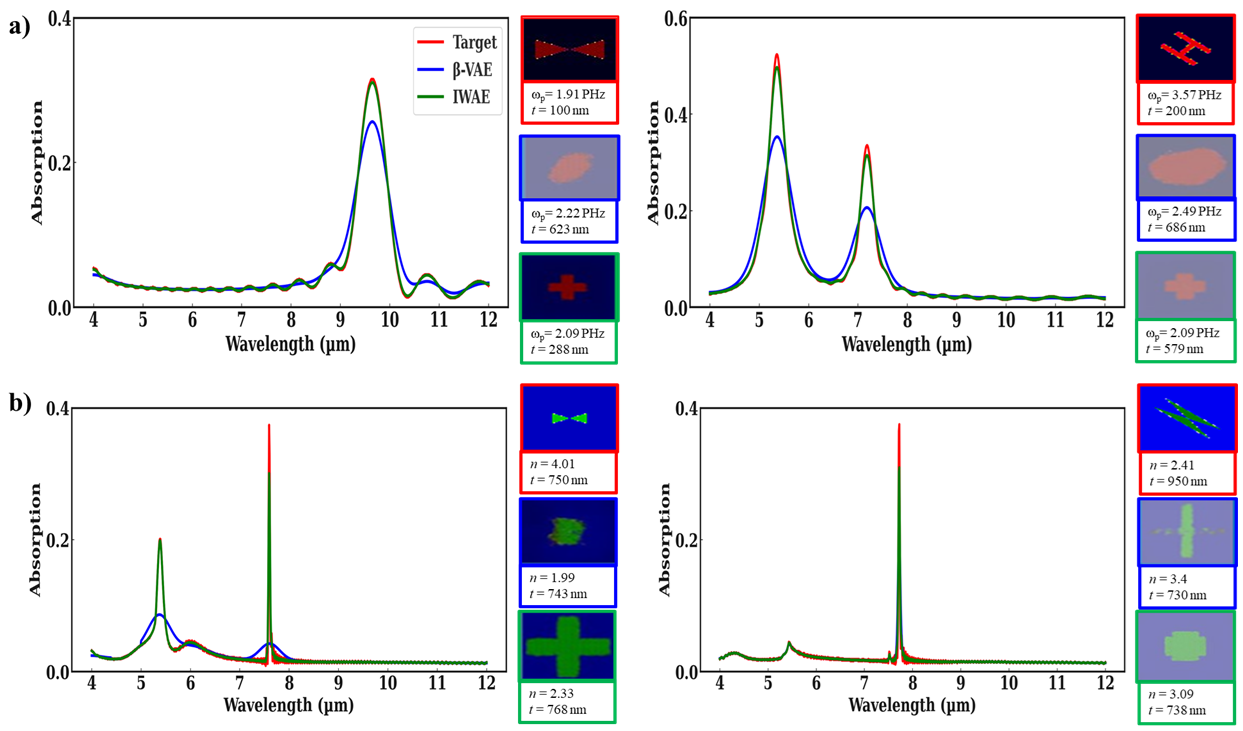}
\caption{Absorption metasurface for both a, b) Lorentzian-shaped and c,d) Fano-shaped spectra from the validation dataset (red), randomly selected as input targets for the La-St QGAN. Simulated spectra from La-St QGAN designs (blue and green) using pretrained \(\beta\)-VAE and IWAE respectively are compared to the targets.}
\label{fig:7}
\end{figure}

The values obtained for plasma frequency (\(\omega_p\)) and refractive index (\(n\)) were utilized to define new materials within the electromagnetic simulations conducted for the design process of the LaSt-QGAN based metasurface. It is crucial to acknowledge that traditional manufacturing techniques may be insufficient for the innovative materials generated through this method.  The framework for material definition enables a broader array of designs influenced by material properties, allowing the model to predict a variety of material characteristics that may be overlooked due to categorical simplifications. To evaluate the effectiveness of the trained La-St QGAN and the applied image processing techniques, a set of Fano parameters was introduced alongside arbitrary latent vectors. In alignment with the established methodology, Figure~\ref{fig:7} presents a series of experiments employing inputs sourced from the validation dataset, which represents 10\% of the training data. The red lines indicate the inputs that have been randomly selected for the study of absorption metasurfaces. The blue and green lines represent the simulated spectra associated with designs produced by La-St QGAN, employing pretrained \(\beta\)-VAE and IWAE methodologies, respectively. The corresponding unit cells are illustrated alongside each plot.

\begin{figure}[ht]
\centering
\includegraphics[width=\linewidth]{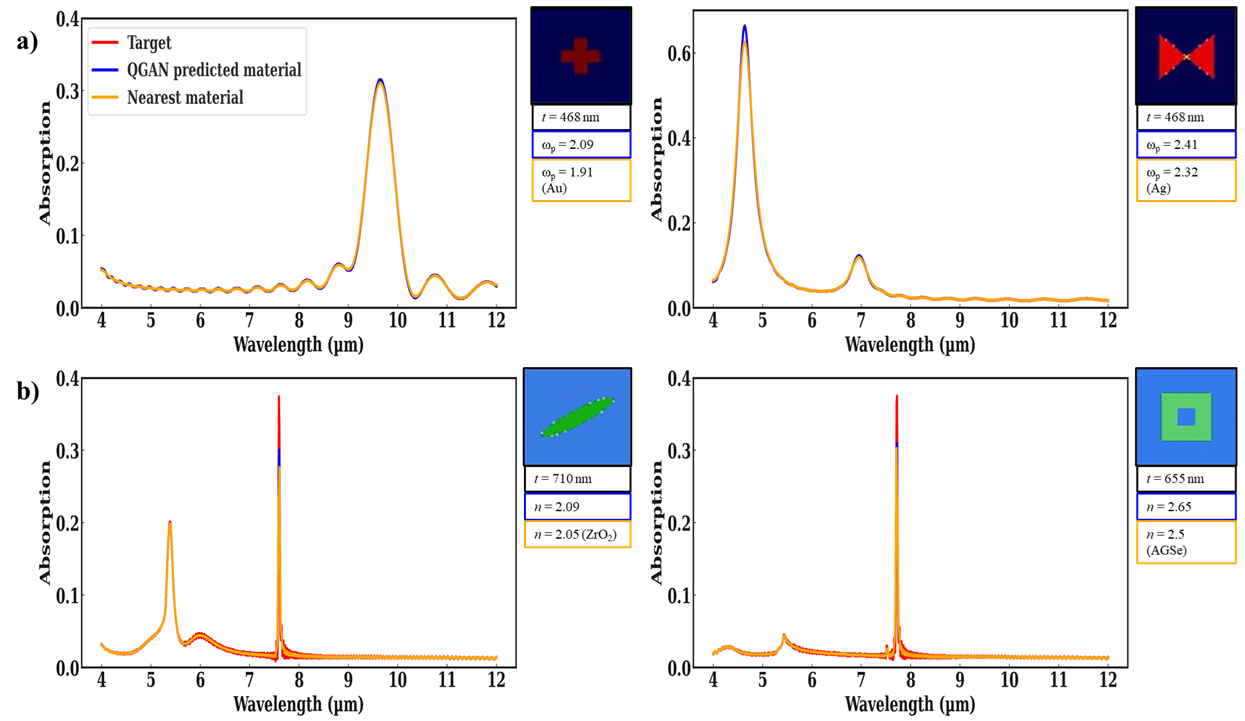}
\caption{
Substitution of LaSt-QGAN-predicted materials with similar alternatives for absorption metasurfaces. a, b) Lorentzian-shaped and c, d) Fano-shaped target absorption spectra (red curves) are compared against simulated outcomes with nearest known materials (orange curves) and original GAN-predicted materials (blue curves).}
\label{fig:8}
\end{figure}

In order to improve the manufacturability and practical applicability of the proposed system, this study introduces the integration of a LaSt-QGAN in conjunction with a publicly accesible database~\cite{polyanskiy2024refractiveindex}. The integration presented herein enables the replacement of estimated material properties with the most closely aligned characteristics derived from conventional materials, as demonstrated in Figure~\ref{fig:8}.  Figure~\ref{fig:8}a and b illustrates the application of this technology on Lorentzian-shaped spectra, where the expected materials are replaced with traditional metals such as silver (Ag) and gold (Au). The re-simulation of these designs revealed that the approximated material properties exhibits 95\% precision in comparison to the results produced by the LaSt-QGAN. The simulated structures exhibit a notable alignment with the anticipated responses, as previously documented in the literature. The application of a pretrained IWAE is employed owing to its proven proximity to the target spectra, as depicted in Figure~\ref{fig:7}.

Figure~\ref{fig:8}c and d presents a series of experiments that focus on input targets characterized by Fano-shaped spectra. The LaSt-QGAN enables the prediction of a range of geometries, with corresponding refraction index values of 2.09 and 2.65, respectively, as one transitions from left to right, illustrating the impact of differing thicknesses. The materials utilized for substitutions included zirconium dioxide (\ce{ZrO2}, \(n\)=2.05) and silver gallium selenide (\ce{AgGaSe2}, \(n\)=2.5). Additionally, it is significant to highlight that certain materials identified through this process, including \ce{ZrO2} and \ce{AgGaSe2}, were absent from the initial training dataset. This highlights the ability of the LaSt-QGAN to forecast new material parameters, including refractive index or plasma frequency, which are essential for identifying alternative materials that satisfy the target requirements, even in cases where these materials were not included in the training dataset with even greater precision.

%The LaSt-QGAN-based methodology demonstrates significant robustness, providing a degree of generalization and design flexibility that exceeds that of conventional machine learning methods, which are limited to making predictions solely within the confines of the training dataset. Conversely, the methodology employed facilitates the investigation of material alternatives that transcend these established boundaries. The scenarios presented illustrate the capability of the LaSt-QGAN to produce predictions that correspond with recognized materials. However, it is acknowledged that the model may also assess properties that extend beyond the conventional boundaries of standard materials. It is anticipated that the accuracy of these predictions will improve as material databases expand a6nd the diversity of available materials increases.

\begin{figure}[ht]
    \centering
    \includegraphics[width=\linewidth]{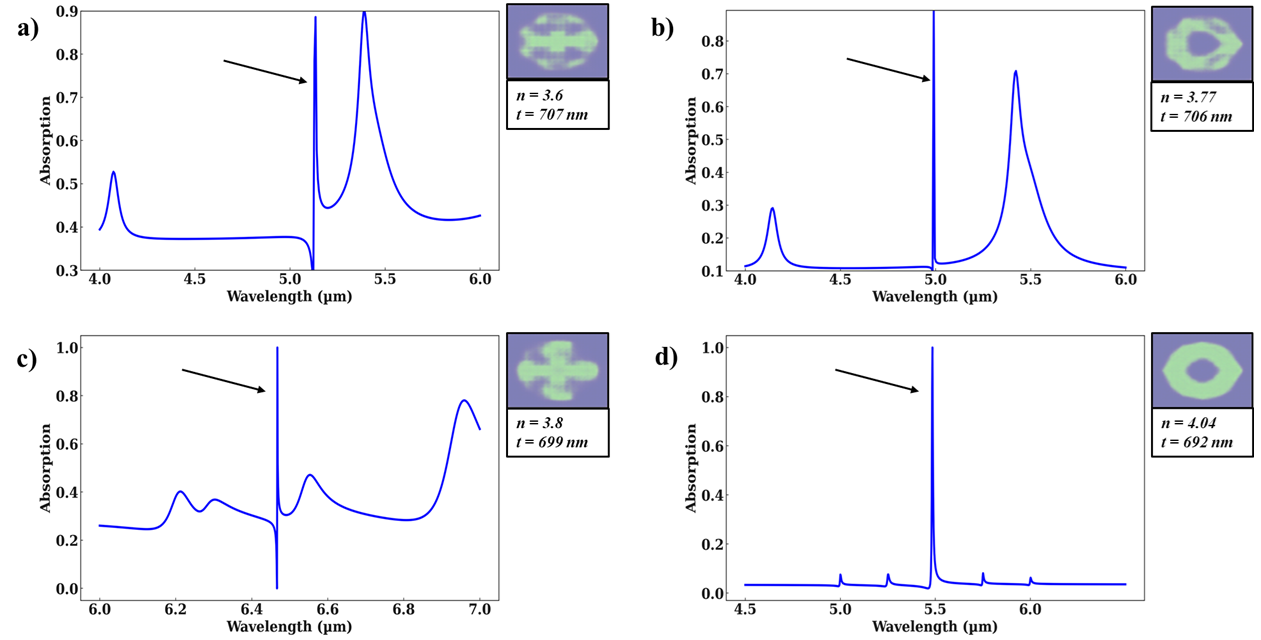}
    \caption{Absorption spectra of metasurfaces exhibiting high-Q Fano resonances at target wavelengths. The inset images show the generated metasurface unit cell designs, which successfully produce the high-Q Fano resonances observed in the spectra.}
    \label{fig:9}
\end{figure}

The absorption spectra of the inversely designed metasurfaces exhibit pronounced high-Q Fano resonances. Panels (a) to (d) correspond to Q-factors of \(1.38 \times 10^4\), \(2.0 \times 10^4\), \(1.2 \times 10^4\), and \(0.95 \times 10^3\), at the target wavelengths of 5.13~\(\mu\)m, 4.99~\(\mu\)m, 6.467~\(\mu\)m, and 5.484~\(\mu\)m, as shown in Figure~\ref{fig:9} respectively. These high-Q resonances result from the engineered interference between discrete resonant modes and the continuum background, producing the characteristic asymmetric Fano line shapes.

Although the inverse design process was based on a dataset primarily composed of symmetric metasurface structures and its corresponding Q-factor values upto the order of $\sim 10^3$, the quantum generative model is capable of generating optimized unit cell geometries exhibiting controlled asymmetry upto the order of $\sim 10^4$. Such asymmetry is a well-established mechanism for achieving high-Q resonances, as it breaks the degeneracy of resonant modes and suppresses radiative losses, thereby sharpening the spectral features.
The inset images depict the corresponding freeform metasurface unit cell designs that produce these high-Q Fano resonances. The emergence of asymmetric geometries from the model highlights its ability to transcend the initial design constraints and explore a broader, more complex parameter space, enabling the discovery of non-intuitive structures that maximize resonance sharpness and spectral selectivity.

\begin{table}[ht]
\centering
\caption{Comparison of conventional GAN and Latent Style Quantum GAN (LaSt-QGAN) performance in inverse design of metasurfaces.}
\label{tab:1}
\begin{tabular}{ccc}
\hline
\textbf{Metric} & \textbf{GAN \cite{Yeung2021global}} & \textbf{LaSt-QGAN} \cr
\hline
Runtime (hrs)     & 25         & 2.5       \cr
MSE               & $10^{-3}$  & $10^{-4}$  \cr
Training Samples & 20,000     & 500        \cr
\hline
\end{tabular}
\end{table}

Table~\ref{tab:1} presents a comparative analysis between a conventional Generative Adversarial Network (GAN), as implemented in Yeung et al.~\cite{Yeung2021global}, and the proposed Quantum GAN (LaSt-QGAN) framework for the inverse design of metasurfaces. The LaSt-QGAN significantly outperforms the classical GAN across all key performance metrics. In terms of runtime, LaSt-QGAN completes training in just 2.5 hours compared to 25 hours for the conventional GAN, demonstrating a tenfold improvement in computational efficiency. Furthermore, the LaSt-QGAN achieves a mean squared error (MSE) of $10^{-4}$, an order of magnitude lower than the GAN’s MSE of $10^{-3}$, indicating higher prediction accuracy. Notably, this improved performance is achieved using only 500 training samples, in contrast to the 20,000 samples required by the conventional GAN, showcasing the LaSt-QGAN’s superior sample efficiency and generalization capability in learning the inverse mapping from target spectra to metasurface designs.

These results underscore the effectiveness of the physics-informed inverse design framework in not only targeting specific resonance wavelengths but also enhancing resonance quality through subtle geometric modifications. This capability is crucial for advancing metasurface functionalities in applications demanding narrowband absorption, enhanced sensing sensitivity, and precise control of electromagnetic wavefronts.

\section{Conclusions}
%The proposed hybrid quantum-classical GAN based inverse design gives rise to metasurface unit cell design with wide-incident angle- independent unidirectional transmission. It is observed that the efficiencies studied in the current work remain independent of angle of incidence in the range $-60^\circ$ to $60^\circ$. The unit cell designed using the proposed framework shows notable enhancements in solar cell performance, particularly in terms of angular efficiency and light absorption. 

The proposed hybrid quantum-classical metasurface design framework demonstrates significant improvements in narrow band absorption. The Latent Style-based Quantum GAN (LaSt-QGAN) is introduced in this study as a hybrid quantum-classical framework for the inverse design of metasurfaces. The model optimizes metasurface structures for narrow-band absorption while also reducing computational costs through the integration of a Quantum Generative Adversarial Network (QGAN) with a Variational Autoencoder (VAE). LaSt-QGAN demonstrates higher fidelity with respect to target absorption spectra compared to the classical GAN. Additionally, it reduces training time by one-tenth and data requirements by 40X when compared to traditional GAN-based methods. The integration of Latent Syle Quantum Generative Adversarial Networks (LaSt-QGANs) streamlines the design process, achieving a 40\% reduction in training time without compromising accuracy.

The framework also addresses the design of high-Q Fano resonances, which enhance light trapping and spectral selectivity. The achieved high-Q resonances play a critical role in boosting metasurface efficiency by optimizing the structural features that govern resonance sharpness and absorption. Additionally, a material look-up table ensures practical manufacturability by replacing predicted materials with real-world alternatives, while maintaining accuracy.

This study highlights the accelerated inverse design methodologies for metasurfaces that can be achieved through quantum-enhanced machine learning. The approach effectively addresses computing inefficiencies and mode collapse in traditional GANs, offering enhanced flexibility in exploring high-dimensional design spaces. The demonstrated improvements underscore the potential of quantum machine learning to revolutionize photonics and materials science, particularly in the optimization of metasurfaces for various applications, including energy harvesting, sensing, imaging, and telecommunications. Future research can further investigate adaptive metasurface designs that dynamically optimize performance under varying environmental conditions, such as changes in spectral and angular distributions of light. Additionally, expanding this approach to incorporate multi-functional metasurfaces and broader photonic applications could drive significant advancements in next-generation optical and energy-harvesting technologies.

\section*{Acknowledgments}
This research is supported and financed by Mahindra University.

\bibliographystyle{unsrt}
\bibliography{references}

\end{document}